\documentclass[reprint,aps,prb,superscriptaddress]{revtex4-1}
\usepackage{natbib}
\usepackage[T1]{fontenc}
\usepackage[latin1]{inputenc}
\usepackage{graphicx, subfigure}
\usepackage{amsmath}
\usepackage{amsthm}
\usepackage{amsfonts}
\usepackage{hyperref}
\begin{document}
\title{Phenomenological theory of the giant magnetoimpedance of composite wires}
\date{\today}
\author{R. Betzholz}
\affiliation{Institute of Experimental Physics, Saarland University, P.O. Box 151150, D-66041 Saarbruecken, Germany}
\affiliation{Department of Physics, East China Normal University, 3663
Zhongshan North Road, 200062 Shanghai, P.R. China}
\author{H. Gao}
\affiliation{Institute of Experimental Physics, Saarland University, P.O. Box 151150, D-66041 Saarbruecken, Germany}
\author{Z. Zhao}
\affiliation{Department of Physics, East China Normal University, 3663
Zhongshan North Road, 200062 Shanghai, P.R. China}
\author{U. Hartmann}
\affiliation{Institute of Experimental Physics, Saarland University, P.O. Box 151150, D-66041 Saarbruecken, Germany}
\begin{abstract}Composite wires with a three-layered structure are known to show a particularly large magnetoimpedance effect. The wires consist of a highly conductive core, an insulating layer and an outer ferromagnetic shell. In order to understand the origin of the effect a theory based on a coupling of the Maxwell equations to the Landau-Lifschitz-Gilbert equation is suggested. The theory is phenomenological in the sense that it does not account for a domain structure. However, theoretical results nicely reproduce those obtained in various measurements. Furthermore, an upper limit of the magnetoimpedance ratio for a given combination of materials can be determined.
\end{abstract}
\maketitle
The giant magnetoimpedance (GMI) effect refers to a huge change in the impedance upon the application of an external static magnetic field $H_0$ to a ferromagnetic conductor. This phenomenon was already mentioned in the 1930s, when the behavior of nickel-iron alloy wires was investigated\cite{Harrison1935,Harrison1936}. In the early 1990s the GMI effect was analyzed more thoroughly for cobalt-based amorphous wires with micrometer dimensions by several groups\cite{Panina1994, Beach1994, Panina1995}. The relative change of the impedance $Z$ is measured in terms of the quantity $\Delta Z/Z$, called the GMI ratio, which is defined as 
\begin{equation}
\frac{\Delta Z}{Z} = \frac{\left| Z(H_0) \right| - \left| Z(H_{ref}) \right|}{\left| Z(H_{ref}) \right|} \ ,
\end{equation}
where $H_{ref}$ represents a reference value of the applied magnetic field. The latter is usually taken to be its maximum value at which the samples can be regarded as magnetically saturated. The ratio can easily reach up to several hundred percent\cite{Vazquez2001} and at the same time exhibit a strong field dependence in the low field regime. This makes the GMI effect promising for applications in sensitive magnetic field detection\cite{Yoon2009,Yu2011}. In order to increase the magnitude of the GMI ratio a wide variety of materials, material combinations and sample geometries have been analyzed\cite{Chen1996,Kraus2003,Zhukova2007,Xiao2000,Zhukova2008}.

Heterogeneous structures, such as layered wires, are of particular interest\cite{Sinnecker2002}. Among them are composite wires consisting of a highly conductive core enclosed by a ferromagnetic shell\cite{Seet2005,Liu2006}. It has been found that the GMI ratio can be further enhanced by inserting an insulating layer between core and shell\cite{Wang2007,Cheng2008,Shi2009}. Ideally, the presence of such an insulating layer restricts the flow of the driving AC current to the inner core only.

In the present work we focus on establishing a model for the calculation and prediction of the GMI ratio in composite wires. The composite wires under investigation are composed of a highly conductive nonmagnetic core, an insulating layer and a ferromagnetic shell. In order to model the impedance of the composite wires the distributions of the electric and magnetic fields are determined in each layer and an expression for the permeability of the ferromagnetic shell is derived. 

The conductive core is characterized by its radius $r_c$ and its axis is along the $z$ axis. When a driving current $I e^{-i \omega t}$ with frequency $f = \omega /2 \pi$ flows through the core, the field distribution is found by solving Maxwell's equations in cylindrical coordinates neglecting displacement currents\cite{Landau1990} and by utilization of Amp\`ere's law as a boundary condition. This restricts the driving current to the core and accounts for a perfect insulation of the inner core from the ferromagnetic shell by the interlayer. The amplitudes $E_z$ of the axial electric field and $H_{\varphi}$ of the circumferential magnetic field are represented by functions that solely depend on the radial coordinate $r$ and can in CGS units be expressed as\cite{Landau1990}

\begin{eqnarray}
E_z(r) &=& C_1 \frac{J_0(k_c r)}{J_1(k_c r_c)} \ , \\ H_{\varphi}(r)  &=& H_I \frac{J_1(k_c r)}{J_1(k_c r_c)} \ ,
\end{eqnarray}

with $H_I = 2 I / c r_c$ and $C_1 = I k_c / 2 \pi \sigma_c r_c$. Here $\sigma_c$ denotes the conductivity of the core material and $c$ the vacuum speed of light. $J_n$ represents Bessel functions of the first kind and order $n$. $k_c$ is defined by $k_c = (1 + i)/\delta_c$, with the skin depth $\delta_c = c/\sqrt{2 \pi \omega \sigma_c}$ of the core.

In the insulating layer of thickness $t_1$ there is no electric current present. The solution of Maxwell's equations for the field amplitudes, which are continuous at the conductor-insulator interface, is then given by

\begin{eqnarray}
E_{z}(r) &=&  C_2  - C_1 k_c r_c \ln \Big( \frac{r}{r_c} \Big) \ , \\ 
H_{\varphi}(r) &=& \frac{2 I}{c r} \ ,
\end{eqnarray}

with $C_2 = C_1 J_0(k_c r_c) / J_1(k_c r_c)$.

The solutions within the ferromagnetic shell, the thickness of which is denoted by $t_2$, are somewhat more involved. Due to $\mathbf{B} = \hat\mu \mathbf{H}$, with the permeability tensor $\hat\mu$, the equations that determine the field amplitudes $H_{\varphi}$ and $H_z$ become now coupled differential equations. As a consequence the solutions of $H_{\varphi}(r)$ and $H_z(r)$ in this region are not Bessel functions. We assume them to be given by series\cite{Buznikov2006} that correspond to the expansions of Bessel functions of the first kind for the decoupled case:

\begin{eqnarray}
H_{\varphi}(r) &=& C_3 \frac {R}{r} + \sum \limits_{n} \Big( a_{n} + c_{n} \ln\Big(\frac{r}{R}\Big) \Big) \Big(\frac{r}{R}\Big)^n \ , \\
H_{z}(r) &=& \sum \limits_{n} \Big( b_{n} + d_{n} \ln\Big(\frac{r}{R}\Big) \Big) \Big(\frac{r}{R}\Big)^n \ , 
\end{eqnarray}

with $R=r_c+t_1+t_2$. Inserting this approach into Maxwell's equations yields recurrence formulae for $a_n$, $b_n$, $c_n$ and $d_n$ and determines the constant $C_3$.  The components of the electric field can be derived from $H_{\varphi}$ and $H_{z}$ by using Maxwell's equations for an ohmic conductor with conductivity $\sigma_m$ ignoring displacement currents: $\nabla \times \mathbf{H} = 4 \pi \sigma_m / c \ \mathbf{E}$. 

In Eqs. (5) and (6), and thereby in the four sets of coefficients, only three combinations of the permeability tensor's elements $\mu_{nm}$ ($n,m = r, \varphi, z$) appear. They are defined by $\mu_{\varphi} = \mu_{\varphi \varphi} - \mu_{r \varphi}^2 / \mu_{r r}$, $\mu_{z} = \mu_{z z} - \mu_{r z}^2 / \mu_{r r}$ and $\mu_{c} = \mu_{\varphi z} - \mu_{r z} \mu_{r \varphi} / \mu_{r r}$. The form of these elements of the permeability tensor have to be derived. This can be achieved by linearizing the Landau-Lifschitz-Gilbert equation\cite{Gilbert2004}
\begin{equation}
\partial_t {\mathbf{M}(t)} = - \gamma {\mathbf{M}(t) \times \mathbf{H}_{eff}} + \frac{\alpha}{M_S} \mathbf{M}(t) \times \partial_t \mathbf{M}(t) \ .
\end{equation}
This equation describes the dynamics of the magnetization vector $\mathbf{M}(t)$ under the influence of the effective magnetic field $\mathbf{H}_{eff}$. $\gamma$ denotes the gyromagnetic ratio of the electron, $\alpha$ the Gilbert damping parameter and $M_S$ the saturation magnetization. Linearization can be achieved by subdividing the magnetization vector $\mathbf{M}(t)$ and the effective magnetic field $\mathbf{H}_{eff}$ into their equilibrium parts $\mathbf{M}_{eq}$ and $\mathbf{H}^{eq}_{eff}$ and into contributions $\mathbf{M} e^{-i \omega t}$ and $\mathbf{H} e^{-i \omega t}$ that are excited by the driving current\cite{Knobel2003}. By using the relation $\mathbf{M}_{eq} \times \mathbf{H}^{eq}_{eff} = 0$ and by omitting second-order terms in the excitations, the Landau-Lifschitz-Gilbert equation can be simplified to\cite{Bertotti1998,Kraus2011}
\begin{equation}
\frac{i \omega}{\gamma} \mathbf{M} = \mathbf{M} \times \Big(\mathbf{H}^{eq}_{eff} + \frac{i \omega \alpha}{\gamma M_S}  \mathbf{M}_{eq}\Big) + \mathbf{M}_{eq} \times \mathbf{H} \ .
\end{equation}
This set of equations establishes a linear relation between the magnetization vector $\mathbf{M}$ and the magnetic field $\mathbf{H}$ through the susceptibility tensor $\hat\chi$. The permeability tensor $\hat\mu$ is then given by $\hat\mu = \hat1+4 \pi \hat\chi$.

The equilibrium part $\mathbf{H}^{eq}_{eff}$ is given by the sum of the external field $\mathbf{H}_0$, which is applied along the $z$ direction, and of a circumferential anisotropy field $\mathbf{H}_K$\cite{Makhnovskiy2001}. Contributions originating from the exchange interaction and the demagnetizing field are not taken into account. The angle $\phi$ between the equilibrium magnetization and the wire axis is derived by minimization of the free energy. This results in
\begin{equation}
H_0 \sin(\phi) - H_K \sin(\phi) \cos(\phi) = 0 \ .
\end{equation} 

The angle is determined by the ratio of the magnitudes of the external field and the anisotropy field strength, $H_0$ and $H_K$. By solving Eq. (9) one is able to obtain

\begin{eqnarray}
\mu_{\varphi} &=& 1 + \mu_{eff} \cos^2(\phi) \ , \\
\mu_{z} &=& 1 + \mu_{eff} \sin^2(\phi) \ , \\
\mu_{c} &=& - \mu_{eff} \sin(\phi) \cos(\phi) \ .
\end{eqnarray}

\begin{figure}[h!]
\centering
 \includegraphics[width=86mm]{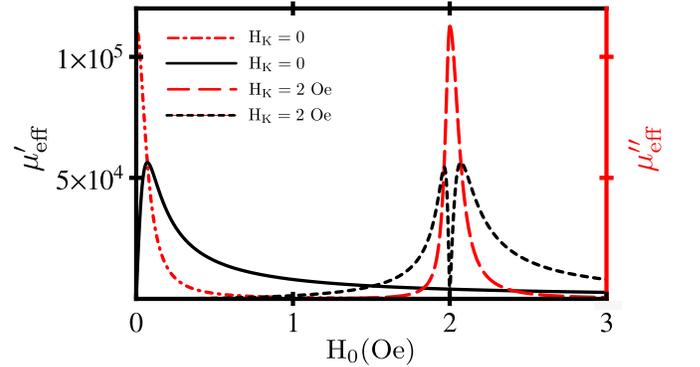}
 \caption{Dependence of the real and imaginary parts of the effective permeability on the applied magnetic field for $H_K = 0$ and $H_K = 2 \ Oe$ at $2$ MHz.}
 \label{1}
\end{figure}

Here, $\mu_{eff}$ denotes an effective permeability. For external field strengths sufficiently far above $H_K$ ($\cos(\phi) \approx 1$) the circumferential permeability $\mu_{\varphi}$ almost equals $\mu_{eff}$. $\mu_{eff}$ consists of a real part $\mu_{eff}'$ and an imaginary part $\mu_{eff}''$ and is defined by
\begin{equation}
\mu_{eff} = \frac{\omega_M (\omega_M + \omega_K - i \omega \alpha)}{(\omega_M + \omega_K - i \omega \alpha) (\omega_K - i \omega \alpha - \omega_0) - \omega^2} \ ,
\end{equation}
with $\omega_M=4 \pi \gamma M_S$, $\omega_K = \gamma H_K$ and $\omega_0 = \gamma H_0 \cos(\phi)$\cite{Usov1998}. As a realistic example we assume $\gamma = 1.76 \cdot 10^7 \ G^{-1}s^{-1}$, $\alpha = 0.1$ and $M_S = 640 \ G$. The field dependence of both the real and the imaginary part of the effective permeability is shown in Fig. \ref{1} for two different values of $H_K$ at a frequency of $2$ MHz. The imaginary part exhibits a sharp peak at $H_0 = H_K$, while the real part shows a double peak in this field regime.
The permeability and field distribution previously derived are used to obtain the impedance $Z$ of the composite wire by consideration of the total energy loss. The input power is given by $P = Z I^2$. On the other hand, the energy dissipated per unit time from a volume element corresponds to the integral of the Poynting vector $\mathbf{S} = c/4 \pi \ \mathbf{E} \times \mathbf{H}$ over the volume's surface. For a composite wire of length $l$ this yields
\begin{equation}
Z = \frac{c l}{2 I^2} \sum\limits_{n = 1}^3 (-1)^n r_n \big(\mathbf{E}(r_n) \times \mathbf{H}(r_n)\big)_r \ ,
\end{equation}
with $r_n$ defining the radial coordinates of the three layers' surfaces. The field amplitudes of $\mathbf{E}$ and $\mathbf{H}$ are obtained from Eqs. (2) - (7). In the ferromagnetic shell the field amplitudes explicitly contain combinations of the permeability tensor's elements with their strong field dependence. Ultimately, the impedance $Z$ is a function of the amplitude $I$ and frequency $f$ of the driving current, of the wire geometry and additionally, through the permeability tensor, of the applied magnetic field and the anisotropy field strength. 

With this expression for the composite wire impedance $Z$ the dependence of the GMI ratio on the external field strength and the frequency of the driving current can be modeled. The GMI ratio exhibits a distinct maximum close to the magnitude of the anisotropy field. This fact is also observed in many experiments\cite{Knobel2003}. Beyond this value $H_K$, the GMI ratio starts to decrease. This yields that in case of a ferromagnetic layer not showing any magnetic anisotropy one would expect a maximum of the GMI ratio at $H_0 = 0$.

\begin{figure}[h!]
\centering
 \includegraphics[width=86mm]{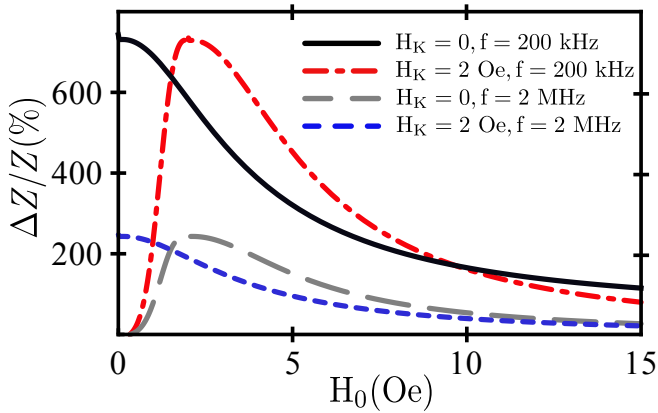}
 \caption{Field dependence of the GMI ratio at $200$ kHz and $2$ MHz for $H_K = 0$ and $H_K = 2 \ Oe$ modeled for a composite wire of a $30$ $\mu$m conductor core, a $3$ $\mu$m insulating layer and a $3$ $\mu$m ferromagnetic shell.}
 \label{2}
\end{figure}

\begin{figure}[h!]
\centering
  \includegraphics[width=86mm]{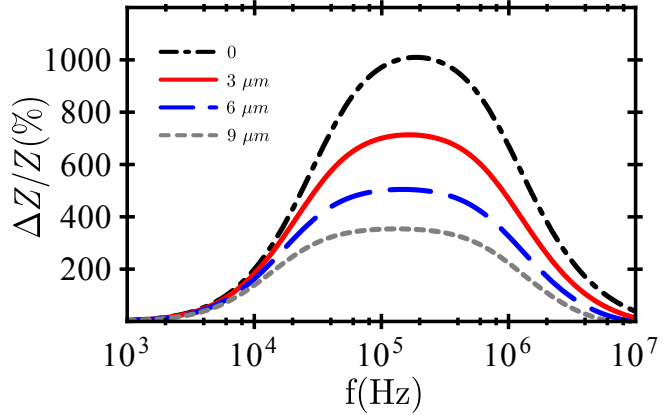}
 \caption{Frequency dependence of the GMI ratio of composite wires with a $30$ $\mu$m core, a $3$ $\mu$m ferromagnetic shell and different thicknesses of the insulating layer.}
 \label{3}
\end{figure}

In the calculations we assume a typical experimental scenario: $r_c = 30 \ \mu m$, $t_1 = t_2 = 0.1 \ r_c$. The remaining parameters are taken to be $\sigma_c = 5 \cdot 10^{17} \ s^{-1}$, $\sigma_m =10^{16} \ s^{-1}$, $H_{ref} = 80 \ Oe$ and $I = 10 \ mA$. Here $\sigma_m$ represents the conductivity of the ferromagnetic shell material. Figure \ref{2} depicts the modeled curves of the field dependence of the GMI ratio for two different values of $H_K$ and two different frequencies of the driving current.
It is obvious that the frequency of the driving current has a large influence on the magnitude of the GMI ratio. The frequency dependence of the GMI ratio at different thicknesses of the insulating layer is shown in Fig. \ref{3}. The maximum of the GMI ratio at $H_0 \approx H_K$ strongly depends on the thickness of the insulating layer with respect to its magnitude but not with respect to the frequency at which it occurs.
The present theory predicts maxima in the $100$ kHz range. This fact is corroborated by experimental results obtained for samples which are comparable to those assumed here\cite{Cheng2008, Zhang2008}. Our results also predict an upper limit of the GMI ratio for $t_1 \rightarrow 0$ and all other parameters as mentioned before. For a vanishing insulator thickness $\Delta Z/Z$ approaches 1000\%. In this case the vanishing insulator layer thickness is assumed to preserve the restriction of the driving current to the core region. This limit expresses that predicted GMI ratios are the highest for insulator layers prepared as thin as possible. If furthermore $t_2 \rightarrow \infty$ is assumed, even 3000\% are reached. These upper limits, which can easily be calculated for any materials combination, provide an important guideline for experimental research.

In conclusion, a model to calculate and predict the GMI ratio of composite wires in a very general form has been established. The impedance of composite wires with insulating interlayer was calculated by separately solving Maxwell's equations in each layer. The solution in the ferromagnetic layer additionally required the derivation of an expression for the permeability tensor. This was achieved by linearization of the Landau-Lifschitz-Gilbert equation which involves several approximations. The field dependence of the GMI ratio for different values of $H_K$ and the frequency dependence for different insulating layer thicknesses have been modeled. For a given frequency the GMI ratio reaches its maximum at an external field strength equal to the value of the anisotropy field. The frequency spectrum indicates large GMI ratios in the $100$ kHz range. The model predicts high GMI ratios for thin insulating and sufficiently thick ferromagnetic layers. The obtained values of 1000\% for very thin insulating layer thicknesses can be compared to experimental results, where maximum GMI ratios of 800\% have been reported\cite{Shi2009}. It is interesting to mention that for other materials theoretical GMI ratio maxima above 3000\% were predicted\cite{Ipatov2010}, which is in the same order of magnitude as our results. 

\end{document}